\documentstyle[tighten,preprint,prl,aps,epsfig]{revtex}
\newcommand {\nc} {\newcommand}
\nc {\beq} {\begin{eqnarray}}
\nc {\eeq} {\end{eqnarray}}
\nc {\siml} [1] {\mathop{\sim}\limits_{r\rightarrow #1}}
\nc {\eeqn} [1] {\label{#1} \end{eqnarray}}
\nc {\ve} [1] {\mbox{\boldmath${#1}$}}
\begin{document}
\draft
\title{Toward a Spin- and Parity-Independent Nucleon-Nucleon Potential}
\author{J.-M.\ Sparenberg}
\address{Physique Nucl\'{e}aire Th\'{e}orique et Physique
Math\'{e}matique, C.P.\ 229,
Universit\'{e} Libre de Bruxelles, B-1050 Brussels, Belgium}
\date{\today}
\maketitle
\begin{abstract}
A supersymmetric inversion method is applied
to the singlet $^1S_0$ and $^1P_1$ neutron-proton elastic phase shifts.
The resulting central potential has a one-pion-exchange (OPE) long-range behavior
and a parity-independent short-range part;
it fits inverted data well.
Adding a regularized OPE tensor term also allows
the reproduction of the triplet $^3P_0$, $^3P_1$ and $^3S_1$ phase shifts
as well as of the deuteron binding energy.
The potential is thus also spin-independent (except for the OPE part) 
and contains no spin-orbit term.
These important simplifications of the neutron-proton interaction are shown
to be possible only if the potential possesses Pauli forbidden bound states,
as proposed in the Moscow nucleon-nucleon model.
\end{abstract}
\pacs{PACS numbers: 13.75.Cs, 03.65.Nk}
\hfill {\em To deep Aline and shallow Clara...} \\[1cm]

Although of fundamental importance for the whole field of nuclear physics
and despite 70 years of intensive research,
a definitive form for the nucleon-nucleon interaction has not yet been established.
Since an individual nucleon is made up of three quarks,
this interaction should in principle be deduced from an {\em ab initio}
quantum chromodynamics description of a six-quark system.
However, such a six-body highly-relativistic problem is out of reach of our
present calculation capacities.
Hence, {\em phenomenological} interactions, 
i.e., interactions constructed to reproduce nucleon-nucleon experimental data
(deuteron bound state and scattering cross sections)
have to be constructed.
These interactions may be divided in two categories.

The first category is based on the theoretical framework of meson exchange 
and fulfills the symmetries of the system.
The best known are the Paris \cite{lacombe:80}, Nijmegen \cite{stoks:94}, Bonn \cite{machleidt:87},
and Argonne \cite{wiringa:95} interactions.
All these models agree on the long-range part of the interaction 
($r > 3$ fm, where $r$ is the inter-nucleon distance):
it is well described by a central (C) and a tensor (T) one-pion-exchange (OPE) potential.
Both terms depend on the total spin $S$ and parity $\pi$ of the nucleon-nucleon system.
The whole potential reads
\beq
V = V^{\mathrm s.r.} + V^{\mathrm OPE}_{\mathrm C}(S,\pi) + V^{\mathrm OPE}_{\mathrm T}(S,\pi),
\eeqn{V}
where the short-range part $V^{\rm s.r.}$ follows 
from the exchange of other mesons and from a purely phenomenological contribution.
Its precise form differs from model to model 
but it at least contains a central, a tensor, a spin-orbit (SO), and a quadratic spin-orbit (QSO) term,
\beq
V^{\mathrm s.r.} = V^{\mathrm s.r.}_{\mathrm C}(E_{\mathrm c.m.}, S, \pi) + 
V^{\mathrm s.r.}_{\mathrm T}(S, \pi) \nonumber \\
+ V^{\mathrm s.r.}_{\mathrm SO}(S, \pi) + 
V^{\mathrm s.r.}_{\mathrm QSO}(S, \pi),
\eeq
all depending on spin and parity.
The central-part dependence on the center-of-mass energy $E_{\mathrm c.m.}$ 
makes the potential non local.

The second category of interactions is physically less justified:
the long-range part is also described by the OPE potential
while the short-range part has a simpler but more phenomenological structure.
In the Reid potential \cite{reid:68}, 
$V^{\mathrm s.r.}$ has neither QSO term nor energy dependence.
The price to pay for this simplification is a partial-wave dependence of the potential:
instead of a dependence on $S$ and $\pi$, 
it depends on $S$ and on the orbital and total angular momenta $L$ and $J$.
There is thus an implicit non-locality and this interaction does not fulfill the symmetries of the system. 
An even simpler form is proposed in the latest version of the Moscow model \cite{kukulin:99}.
In this case, the short-range potential consists of a local part only depending on spin and parity,
without tensor term,
\beq
V^{\mathrm s.r.}_{\mathrm local} = V^{\mathrm s.r.}_{\mathrm C}(S, \pi) + V^{\mathrm s.r.}_{\mathrm SO}(S, \pi),
\eeqn{mosc}
and a non-local separable term depending on the partial wave $V^{\mathrm s.r.}_{\mathrm sep}(S,L,J)$.
This further simplification of the interaction structure is made possible by the introduction of
Pauli forbidden bound states (PFSs) in the lowest partial waves.
These non-physical states are aiming to simulate the Pauli exclusion principle between the six quarks
constituting the interacting nucleons \cite{kukulin:92}.
Their presence results in a {\em deep} potential without repulsive core,
whereas other nucleon-nucleon interactions, 
which only possess {\em one} physical bound state (the deuteron),
are {\em shallow} and present a repulsive core below about 1 fm.

In the present work,
we show that the use of a deep potential allows a further simplification of the 
short-range local part (\ref{mosc}).
Neither the spin and parity dependence nor the spin-orbit term seem to be necessary 
to fit data, so that one only has
\beq
V^{\mathrm s.r.}_{\mathrm local}=V^{\mathrm s.r.}_{\mathrm C}.
\eeqn{Vsr}
Spin and parity dependences,
or equivalently spin-spin and isospin-isospin terms,
are generally considered as essential ingredients of the nucleon-nucleon interaction,
in contrast with Eq.\ (\ref{Vsr}).
Besides the presence of PFSs, 
the present simplification requires a more complicated central part than
in the Moscow model, 
where an exponential or Gaussian form factor is assumed from the beginning.
Here we deduce this form factor from neutron-proton scattering data through the use
of the supersymmetric inversion method of Ref.\ \cite{sparenberg:00c}.
This technique allows the construction of a partial-wave-independent potential from
elastic scattering phase shifts of several partial waves.
In Ref.\ \cite{sparenberg:00c}, 
the inversion method is applied to the system of two $\alpha$ particles,
for which the inclusion of PFSs has been known to considerably simplify the interaction for
a long time.
We show here that a similar simplification occurs in the interaction between two nucleons,
a fact ignored up to now.

\begin{figure}
\begin{center}
\scalebox{0.5}{\includegraphics[70,280][540,550]{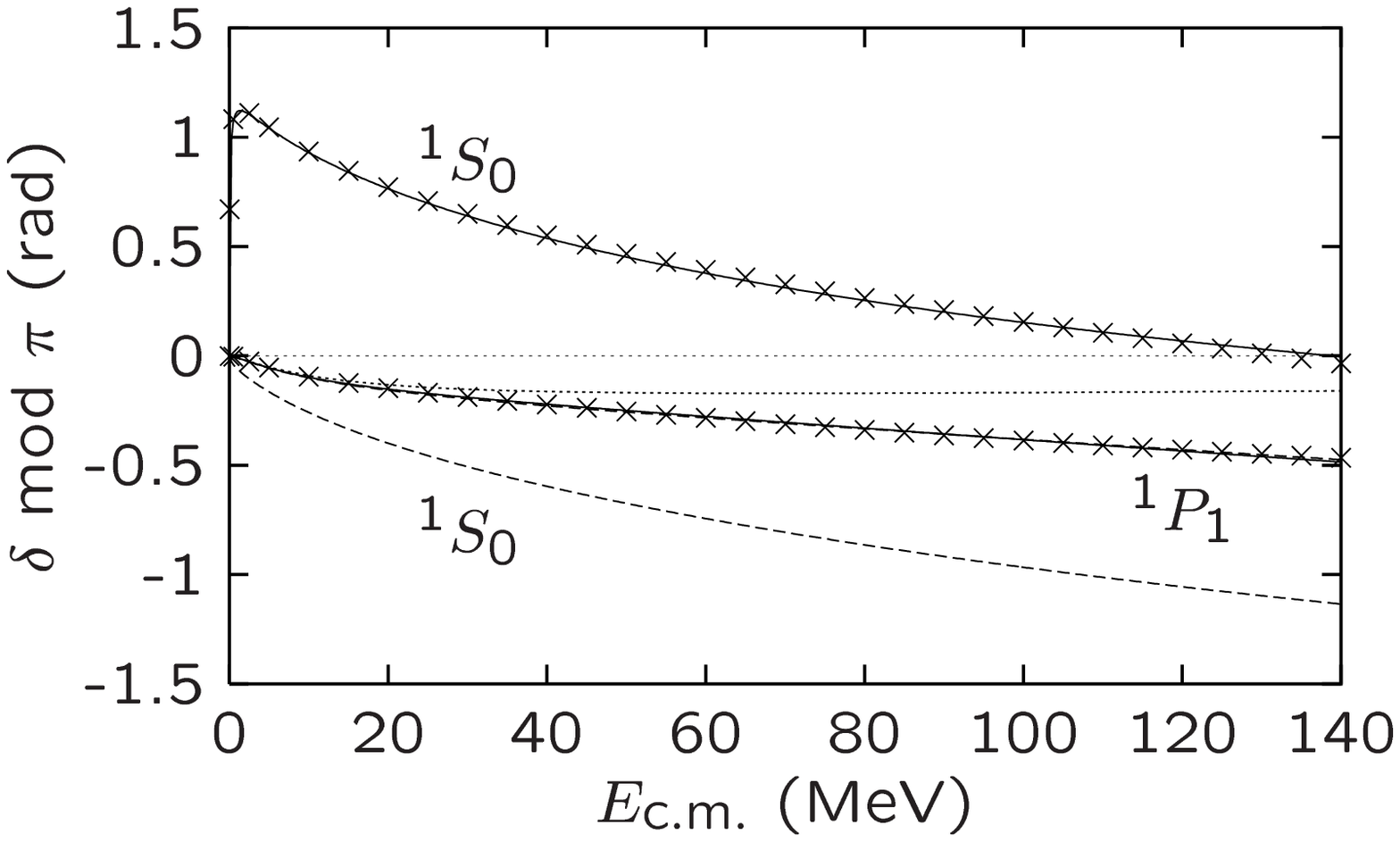}}
\end{center}
\caption{\label{d01} Neutron-proton $^1S_0$ and $^1P_1$ phase shifts:
data from Ref.\ \protect\cite{stoks:93} (crosses), shallow- (dashed lines) and deep-potential (solid lines)
calculations. The $^1P_1$ phase shifts of both potentials are almost indistinguishable.
The $^1P_1$ phase shifts of the OPE potential (\ref{OPEc}) for $\Lambda_{\mathrm C}=468.1$ MeV/$\hbar c$ 
(dotted line) are also shown.}
\end{figure}

Our inversion method is a two-step procedure \cite{sparenberg:00c}.
We first invert the phase shifts $\delta(k)$ of the singlet ($S=0$) $^1P_1$ partial wave.
As ``experimental'' data
we use the multienergy phase shifts PWA93 of Ref.\ \cite{stoks:93} at 
$E_{\mathrm c.m.}=0.05, 0.5, 2.5, 5, 10, 15, \dots, 140$ MeV 
(below the inelastic pion-production threshold).
The wave number $k$ (in fm$^{-1}$) is related to $E_{\mathrm c.m.}$ (in MeV) through
$k=$ $\sqrt{2 \mu E_{\mathrm c.m.}/\hbar^2}$,
where $\mu = 469.45926$ MeV/$c^2$ is the neutron-proton reduced mass with
$\hbar c=197.327053$ MeV fm.
We use a rational approximation of the scattering $S$-matrix
\beq
S(k) \equiv e^{2\imath\delta(k)}
\approx S^{\mathrm OPE}_{\mathrm C} (k) \prod_{m=1}^M \frac{\kappa_m+k}{\kappa_m-k},
\eeqn{Srat}
where $S^{\mathrm OPE}_{\mathrm C}$ is the $S$-matrix of $V_{\mathrm C}^{\mathrm OPE}$
($V_{\mathrm T}^{\mathrm OPE}$ vanishes for singlet waves)
and $\kappa_m$ are complex wave numbers fitted to experimental data.
Potential $V_{\mathrm C}^{\mathrm OPE}$ depends on spin and parity through the spin and isospin
Pauli matrices of the nucleons $\ve{\sigma_1}$, $\ve{\sigma_2}$, $\ve{\tau_1}$, $\ve{\tau_2}$.
It reads (in MeV)
\beq
V^{\mathrm OPE}_{\mathrm C}(S,\pi,r) = \frac{\ve{\tau_1} \cdot \ve{\tau_2}}{3}\; 0.075 \;
m_\pi c^2\; \Phi^0_{\mathrm C}(r) \, (\ve{\sigma_1} \cdot \ve{\sigma_2}),
\eeqn{OPEc}
where an averaged pion mass $m_\pi=138.0363$ MeV/$c^2$ has been used
(distinguishing between neutral- and charged-pion exchange does not qualitatively
modify the present results).
A dipole-regularized form factor $\Phi^0_{\mathrm C}$ is used [Eq.\ (9) of Ref.\ \cite{stoks:94}]
with a cutoff parameter $\Lambda_{\mathrm C}$ given below.
Other form-factor choices (as in Ref.\ \cite{kukulin:99}) will be discussed elsewhere.
For the $^1P_1$ partial wave, 
one has $\ve{\tau_1} \cdot \ve{\tau_2} = \ve{\sigma_1} \cdot \ve{\sigma_2} = -3$ in Eq.\ (\ref{OPEc}).
The corresponding OPE phase shifts are represented by a dotted line in Fig.\ \ref{d01}.
They are satisfactory at low energy only.
With three additional $S$-matrix poles [$M=3$ in Eq.\ (\ref{Srat})] located at
$\kappa_m=\pm1.848+ 1.860 \imath$ and $-1.848 \imath$ fm$^{-1}$,
a perfect data fit is obtained on the whole energy range 
(upper dashed line in Fig.\ \ref{d01}).
The values of the poles have been calculated with the method presented in Ref.\ \cite{sparenberg:97a}.

\begin{figure}
\begin{center}
\scalebox{0.48}{\includegraphics[30,280][530,550]{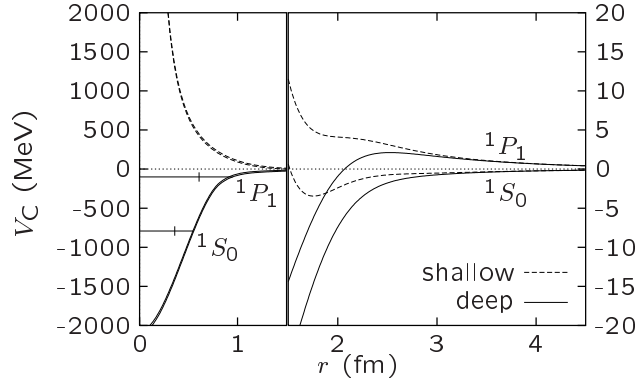}}
\end{center}
\caption{\label{p01} Shallow (dashed lines) and deep (solid lines) central potentials
obtained by supersymmetric inversion for the $^1S_0$ and $^1P_1$ partial waves.
The $^1S_0$ and $^1P_1$ potentials are almost indistinguishable in the left panel,
where the energy (horizontal lines) and mean radius (vertical lines) 
of the deep-potential PFSs are also shown for both partial waves.}
\end{figure}

The potential corresponding to the rational $S$-matrix (\ref{Srat}) is a generalized Bargmann potential
\cite{chadan:77}.
It may be constructed by three successive supersymmetric transformations \cite{sukumar:85}
of the $^1P_1$ radial Schr\"odinger equation with the OPE potential.
It has no bound state and is purely repulsive (upper dashed line in Fig.\ \ref{p01})
in agreement with traditional shallow nucleon-nucleon interactions.
To calculate $^1S_0$ phase shifts,
we modify the OPE term with $\ve{\tau_1} \cdot \ve{\tau_2}= 1$ instead of $-3$ in Eq.\ (\ref{OPEc})
and we keep the same short-range part [Eqs.\ (\ref{V}) and (\ref{Vsr})].
This makes the central potential slightly attractive above 1.5 fm 
(lower dashed line in Fig.\ \ref{p01}).
However, the repulsive $V^{\mathrm s.r.}_{\mathrm C}$ dominates and the $^1S_0$ phase shifts are negative
(lower dashed line in Fig.\ \ref{d01}),
in disagreement with experimental data.
Modifying $\Lambda_{\mathrm C}$ does not solve the problem.
This explains why shallow nucleon-nucleon potentials always have a strong parity dependence:
a shallow potential is unable to fit both the $^1S_0$ and $^1P_1$ phase shifts simultaneously.

Let us now perform the second step of our inversion technique \cite{sparenberg:00c},
namely, the addition of a Pauli forbidden bound state to the $^1P_1$ effective potential
just obtained,
in order to construct a deep potential.
Such an addition may be done without modifying the $^1P_1$ phase shifts 
(phase-equivalent addition) with the help of two successive supersymmetric transformations
\cite{baye:87b}.
Both steps of the inversion method are thus performed with the
same algebraic formalism of supersymmetric transformations.
Let us define the mean radius $r_{\mathrm PFS}^{\mathrm mean}$ of the added PFS by
\beq
\int_{r_{\mathrm PFS}^{\mathrm mean}}^\infty |\psi_{\mathrm PFS}(r)|^2 dr =\frac{1}{2},
\eeq
where $\psi_{\mathrm PFS}$ is the PFS normalized wave function.
An important feature of the phase-equivalent PFS addition is that both 
the binding energy $E_{\mathrm PFS}$ and $r_{\mathrm PFS}^{\mathrm mean}$ may be chosen arbitrarily.
Here we choose them to fit the $^1S_0$ phase shifts together with the $^1P_1$ ones.
However, the obtained central potential has a singular $r^{-2}$ attractive core at the origin 
and cannot be used as such to estimate $^1S_0$ phase shifts.
Hence we regularize it as $V_{\rm reg}+ a r^2 + b r^3$ below a regularization radius $r^{\rm reg}$,
as in Ref.\ \cite{sparenberg:00c}.
This only slightly affects the $^1P_1$ phase shifts provided $r_{\rm reg}$ is lower than the classical
turning point for this partial wave at the highest considered energy.

For $E_{\mathrm PFS}=-101.9$ MeV, $r^{\mathrm mean}_{\mathrm PFS}=0.6064$ fm,
$V_{\mathrm reg}=-2071$ MeV, $r_{\mathrm reg}=0.8231$ fm and $\Lambda_{\mathrm C}=468.1$ MeV/$\hbar c$,
we get a very good fit of the $^1S_0$ and $^1P_1$ phase shifts (solid lines in Fig.\ \ref{d01}),
with root-mean-square (rms) deviations of 3.9 and $2.3 \times 10^{-3} \pi$ respectively .
The corresponding central potentials are deep and have no repulsive core.
They are represented by solid lines in Fig.\ \ref{p01},
together with the energy and mean radius of their PFSs.
Let us recall that, by construction, the s.r.\ term of the potential is parity independent
since it fits both an odd ($^1P_1$) and an even ($^1S_0$) wave.
Thanks to the soft cutoff parameter,
the OPE term only introduces a small parity dependence.
Such a soft cutoff is shown to be very satisfactory in Ref.\ \cite{kukulin:99}
and seems to be a constant feature of deep potentials.

\begin{figure}
\begin{center}
\scalebox{0.5}{\includegraphics[60,280][530,550]{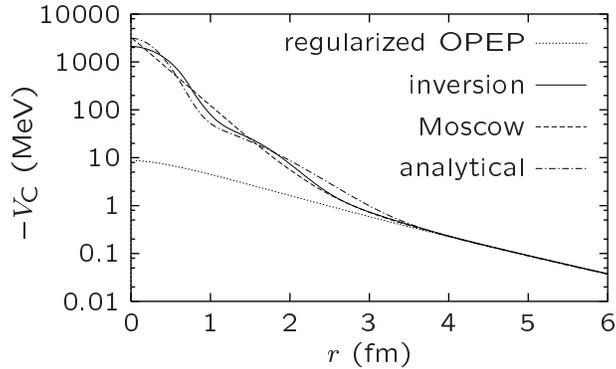}}
\end{center}
\caption{\label{p1S0} Central potentials for the $^1S_0$ partial wave:
dipole-regularized OPE potential (\ref{OPEc}) with $\Lambda_{\mathrm C}=468.1$ MeV/$\hbar c$
(dotted line), deep inversion potential (solid line), deep Moscow potential of Ref.\
\protect\cite{kukulin:99} (dashed line), sum of the analytical s.r.\ potential (\ref{Vanal}) and
of the OPE potential (\ref{OPEc}) with $\Lambda_{\mathrm C}=406.2$ MeV/$\hbar c$ (dash-dotted line).}
\end{figure}

In Fig.\ \ref{p1S0}, we compare the form factor of our potential (solid line) 
with that of the local part of the Moscow potential \cite{kukulin:99} (dashed line)
for the $^1S_0$ partial wave.
Both potentials have an OPE tail (dotted line) but
have different short-range behaviors.
In the Moscow model, an exponential form factor is used,
while the inversion potential has a more complicated form factor,
which could not have been easily guessed from phase shifts.
The inversion potential has no compact analytical expression but
it is reasonably well approximated by a sum of two Gaussians
\beq
V^{\mathrm s.r.}_{\mathrm C}(r)=-3050\, e^{-5.890 r^2}-70.39\, e^{-0.5687 r^2}
\eeqn{Vanal}
(in MeV, $r$ in fm),
as shown by the dash-dotted line in Fig.\ \ref{p1S0}.
Adding an OPE term with $\Lambda_{\mathrm C}=406.2$ MeV/$\hbar c$ to potential (\ref{Vanal})
provides rms deviations of 9.3 and $2.4 \times 10^{-3} \pi$ for the 
$^1S_0$ and $^1P_1$ phase shifts respectively,
which is still in good agreement with experimental data.
In the following, we only use the numerical inversion potential but we have checked that
our conclusions also hold for the analytical potential (\ref{Vanal}),
which is of easier use for practical applications.

\begin{figure}
\begin{center}
\scalebox{0.5}{\includegraphics[70,280][540,550]{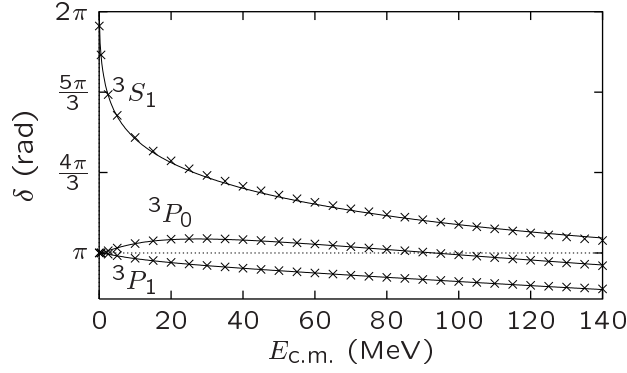}}
\end{center}
\caption{\label{dt01} Neutron-proton $^3S_1$, $^3P_0$ and $^3P_1$ phase shifts:
data from Ref.\ \protect\cite{stoks:93} (crosses) and calculations (solid lines)
with the central inversion potential and an OPE tensor term.}
\end{figure}

Let us now show that the inversion potential is also able to fit triplet ($S=1$) partial waves,
in particular the deuteron bound state in the $^3S_1$-$^3D_1$ coupled waves.
We have found that it is sufficient to add a tensor OPE term to the above central potential
to reproduce the experimental deuteron binding energy $E_{\mathrm d}=-2.224575(9)$ MeV
as well as the $^3S_1$ phase shifts (Fig.\ \ref{dt01}) 
with an rms deviation of $6.0 \times 10^{-3} \pi$.
A dipole-regularized form factor $\Phi^0_{\mathrm T}$ [Eq.\ (12) of Ref.\ \cite{stoks:94}]
is used with a smooth cutoff parameter $\Lambda^{\mathrm even}_{\mathrm T}=601.14$ MeV/$\hbar c$.
The deuteron wave function is then similar to that of the Moscow
potential (Fig.\ 5 of Ref.\ \cite{kukulin:99}) and has a node in the $S$ wave.
The triplet-odd $^3P_0$ and $^3P_1$ phase shifts are also reproduced 
(Fig.\ \ref{dt01}, rms deviations of 1.3 and $0.9 \times 10^{-3} \pi$ respectively)
with a cutoff parameter $\Lambda^{\mathrm odd}_{\mathrm T}=517.7$ MeV/$\hbar c$.
Finally, the mixing parameters have a good qualitative behavior
but $\epsilon_1$, $\epsilon_2$ and $\epsilon_3$ are too small;
this indicates that the tensor form factor could be slightly improved.

The parity-independent central potential deduced from inversion of 
singlet-even and singlet-odd phase shifts
is thus also valid for triplet-even and triplet-odd partial waves.
Hence in addition to the parity, it is independent of the spin.
To our knowledge,
the present potential is the simplest phenomenological model of the 
nucleon-nucleon interaction able to fit all the above phase shifts simultaneously.
Let us insist on the fact that the potential is purely local 
(no energy dependence, no $J$ or $L$ dependence) and has no spin-orbit term.
The potential of Ref.\ \cite{kukulin:99}, on the contrary,
has a spin-orbit and a non-local term for the $^3P_0$ and $^3P_1$ waves.
Moreover, it has no PFS for these waves,
whereas our potential has one PFS for all $S$ and $P$ waves.
The present model has thus a simpler structure.

\begin{figure}
\begin{center}
\scalebox{0.5}{\includegraphics[70,280][540,550]{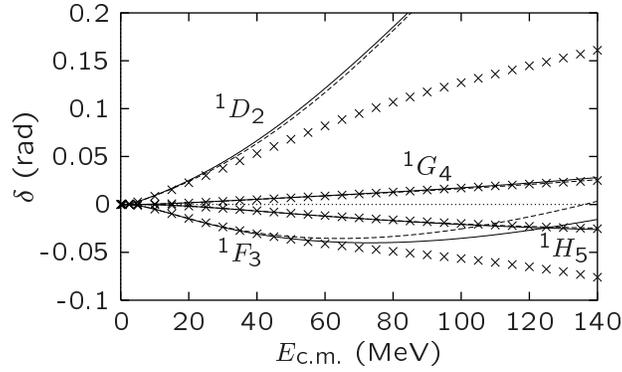}}
\end{center}
\caption{\label{dsinglet} Neutron-proton singlet phase shifts:
data from Ref.\ \protect\cite{stoks:93} (crosses), 
calculations with the inversion potential (solid lines)
and with the local part (\ref{mosc}) of the potential of Ref.\ \protect\cite{kukulin:99}
(dashed lines).}
\end{figure}

However, our potential does not reproduce the phase shifts of every higher 
partial waves ($J \ge 2$),
which means that the present work is only a first step toward a simpler
nucleon-nucleon interaction.
In particular, we get negative $^3P_2$ phase shifts
whereas the Moscow model agrees with the positive experimental values.
Moreover, the $D$ and $F$ phase shifts have a satisfactory low-energy behavior
but are too large at higher energies,
as illustrated for instance in Fig.\ \ref{dsinglet} (solid lines) for the singlet partial waves.
A similar result is obtained for triplet waves.
The higher partial waves ($G$, $H$, \dots) are well reproduced because they mainly
depend on the OPE tail of the potential.
The discrepancy for the $D$ and $F$ waves is also encountered 
with the local part (\ref{mosc}) of the Moscow potential
(dashed lines in Fig.\ \ref{dsinglet}).
In Ref.\ \cite{kukulin:99}, this is solved by the introduction 
of a repulsive pseudo-potential term $V_{\mathrm sep}^{\mathrm s.r.}$ for these
partial waves.
More conventional non-local terms, 
such as an energy-dependent central part,
could also be used and the possibility of a spin- and parity-independent 
non-local term could be considered.

As a conclusion, this work gives some hope for an important simplification 
of the phenomenological nucleon-nucleon interaction.
A satisfactory local part has been found,
which consists of a deep short-range central potential,
independent of both spin and parity,
and a regularized OPE potential (central and tensor terms).
A spin-orbit term does not seem to be necessary,
while the question of a satisfactory non-local term is still open.
We plan to apply the same technique to proton-proton data
(the Coulomb interaction does not complicate the inversion method \cite{sparenberg:97a})
in order to study the charge dependence of our model.
We would also like to generalize the present inversion technique to the coupled-channel case,
which would allow a deduction of the tensor form factor from experimental data.
First steps in this direction have already been made \cite{sparenberg:97b,leeb:00}.

I thank Profs.\ V.\ I.\ Kukulin and V.\ N.\ Pomerantsev for information on
their OPE form factor,
as well as Prof.\ D.\ Baye for his careful reading of the manuscript and
stimulating discussions.
I am supported by the National Fund for Scientific Research, Belgium.
This text presents research results of the Belgian program P4/18 on inter-university 
attraction poles initiated by the Belgian-state Federal Services for
Scientific, Technical and Cultural Affairs.


%

\end{document}